\documentclass[apl,reprint,superscriptaddress,super,superbib,amsmath,aip]{revtex4-1}
\usepackage{graphicx}
\usepackage{color}
\usepackage{ulem}
\usepackage{siunitx}

\begin{document}   

\title{Resonance fluorescence of a site-controlled quantum dot realized by the buried-stressor growth technique}

\author{Max Strau\ss} 
\affiliation{Institut f\"ur Festk\"orperphysik, Technische Universit\"at Berlin,
 10623 Berlin, Germany}
\author{Arsenty Kaganskiy}
\affiliation{Institut f\"ur Festk\"orperphysik, Technische Universit\"at Berlin,
 10623 Berlin, Germany}
\author{Robert Voigt}
\affiliation{Institut f\"ur Festk\"orperphysik, Technische Universit\"at Berlin,
 10623 Berlin, Germany}
\author{Peter Schnauber}
\affiliation{Institut f\"ur Festk\"orperphysik, Technische Universit\"at Berlin,
 10623 Berlin, Germany}
\author{Jan-Hindrik Schulze}
\affiliation{Institut f\"ur Festk\"orperphysik, Technische Universit\"at Berlin,
 10623 Berlin, Germany}
\author{Sven Rodt}
\affiliation{Institut f\"ur Festk\"orperphysik, Technische Universit\"at Berlin,
 10623 Berlin, Germany}
\author{Andre Strittmatter}
\affiliation{Institut f\"ur Festk\"orperphysik, Technische Universit\"at Berlin,
 10623 Berlin, Germany}
\affiliation{Present address: Abteilung f\"ur Halbleiterepitaxie, Otto-von-Guericke Universit\"at, 39106 Magdeburg,Germany}
\author{Stephan Reitzenstein }
\email{stephan.reitzenstein@physik.tu-berlin.de}
\affiliation{Institut f\"ur Festk\"orperphysik, Technische Universit\"at Berlin,
 10623 Berlin, Germany}

\begin{abstract}
Site-controlled growth of semiconductor quantum dots (QDs) represents a major advancement to achieve scalable quantum technology platforms. One immediate benefit is the deterministic integration of quantum emitters into optical microcavities. However, site-controlled growth of QDs is usually achieved at the cost of reduced optical quality. Here, we show that the buried-stressor
growth technique enables the realization of high-quality site-controlled QDs with attractive optical and quantum optical properties. This is evidenced by performing excitation power dependent resonance fluorescence experiments at cryogenic temperatures showing QD emission linewidths down to 10 $\mu$eV. Resonant excitation leads to the observation of the Mollow triplet under CW excitation and enables coherent state preparation under pulsed excitation. Under resonant $\pi$-pulse excitation we observe clean single photon emission associated with $g^{(2)}(0)=0.12$ limited by non-ideal laser suppression.
\end{abstract}

\maketitle   
Quantum emitters are central objects in emerging quantum technologies~\cite{Kimble2008}. Quantum communication, for instance, is based on single photons as information carriers~\cite{Gisin2007}. While simple quantum key distribution 
can be implemented by strongly attenuated lasers, advanced schemes for long distance quantum communication require entanglement distribution which requires real quantum emitters with a non-classical 
photon statistics, high photon indistinguishably and high extraction efficiency~\cite{Sangouard2007}. Recent experiments have shown that self-assembled semiconductor quantum dots (QDs) are prime candidates to meet these
 requirements~\cite{Ding2016}. Moreover, in contrast to non-classical light sources relying on spontaneous parametric down conversion~\cite{Kwiat1995}, QDs offer the great prospect of providing single photons on 
 demand~\cite{Ding2016}. The down-side of growing QDs by self-assembly is randomness in position and emission energy. This is particularly problematic when it comes to device integration. As a result, in-situ lithography techniques were invented to circumvent this issue by pre-selecting suitable QDs from a large ensemble~\cite{Dousse2008,Gschrey2013}. On the other hand, on-chip schemes for photonic computing are 
 usually based on regular arrays of coupled single photon emitters, or on quantum emitters integrated into regular waveguides \cite{Laucht2012, Makhonin2014}.
 
In order to facilitate scalable device concepts based on QDs, different schemes for their site-controlled growth have been developed. A prominent example applies arrays of etched nanoholes and inverted pyramids as 
nucleation centers for the localized growth of QDs \cite{Pelucchi2007,Surrente2009,Schneider2009a,Pfau2009}. This
approach leads to excellent site-control of the QDs position and allows for the device integration of spatially aligned single QDs.
However, tight site-control goes along with enhanced impact of defect centers and non-radiative recombination when the QDs form in close proximity
 to the etched nanoholes \cite{Albert2010}. Indeed, 
there is a general trade-off between site-selectivity and optical quality of site-controlled quantum dots (SCQDs)
\cite{Jons2013,Unsleber2015a}. Other approaches based on SCQDs in non-planar sample geometries such as QDs embedded in nanowires have shown promise regarding the optical quality of the QDs \cite{Huber2014} but offer challenges regarding their scalability and integrability. Another promising technology platform for the realization of site-controlled QDs in planar sample geometries is based on a buried stressor~\cite{Unrau2012,Strittmatter2012b}. In this approach, 
strain-tuning by a oxide-aperture leads to the localized formation of QDs where the number of site-controlled QDs can be controlled by the diameter of the aperture\cite{Strittmatter2012b,Strittmatter2012}. The approach provides a pristine growth surface fairly separated from the stressor and, thus, promises high optical quality of the localized QDs.

In this letter, we report on optical and quantum optical properties of single
QDs grown on a buried-stressor under strict resonant excitation. We observe the
Mollow triplet under cw-excitation allowing us to identify exciton-phonon
coupling as dominant dephasing mechanism in our system. Furthermore, by using
resonant pulsed excitation to deterministically prepare inversion of our
two-level system, we are able to demonstrate single photon emission 
($g^{(2)}(0)=0.12$).

\begin{figure}[htbp]
\includegraphics[width=\linewidth]{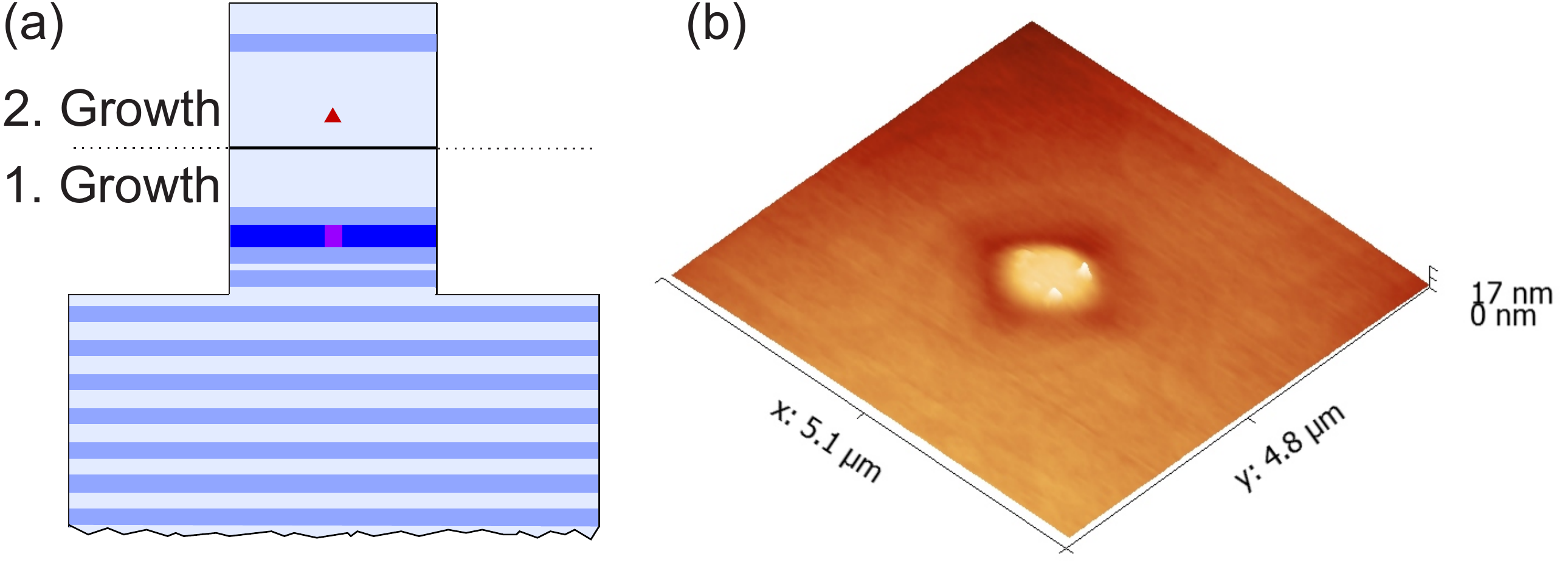}
\caption{\label{graph:0} (a) Schematic view of a fully processed and overgrown structure with SCQD. (b) Atomic force microscopy (AFM) image of two site-controlled quantum dots positioned over a buried 
stressor with an aperture diameter of $\approx$~700~nm.}
\end{figure}

The SCQDs were grown via metal-organic chemical vapor deposition (MOCVD) using an n-doped GaAs substrate. The fabrication process starts with the growth of a template for subsequent
 etching of mesa-structures and oxidation of local apertures \cite{Strittmatter2012b}. The layer structure comprises a distributed Bragg reflector (DBR) mirror consisting of 27~pairs of $\lambda$/4 thick 
 $\textrm{Al}_{\textrm{0.90}}\textrm{Ga}_{\textrm{0.10}}\textrm{As}/\textrm{GaAs}$ followed by a 30~nm thick $\textrm{AlAs}$ layer embedded into 40~nm thick
 $\textrm{Al}\textrm{Ga}\textrm{As}$ claddings. Following the initial growth step, 20 to 21~$\mu$m wide and square shaped mesas are formed via reactive ion etching in a inductively coupled 
 plasma (ICP-RIE). The square shape of the mesas is chosen as to minimise the formation of defects during overgrowth which tend to form faster on side walls in [110] and [1-10] directions. This etching removes the semiconductor material down to the lowest DBR mirror pair to laterally expose the $\textrm{AlAs}$ layer for oxidation performed at 420~$^{\circ}$C in a $\textrm{H}_{\textrm{2}}\textrm{O}/\textrm{N}_{\textrm{2}}$ atmosphere. Here, in-situ optical monitoring of 
 the oxidation process facilitates controlling the aperture diameter on the scale of a few hundred nanometers. Afterwards, the oxidized mesas are overgrown by a 50~nm thick GaAs 
 buffer layer followed by the site-controlled InGaAs QDs sandwiched between the oxide aperture and an upper $\textrm{AlGaAs}$ charge-carrier antidiffusion barrier and a thin (30~nm) $\textrm{GaAs}$ capping layer (cf. Fig.~\ref{graph:0}~(a)).

Our site-controlled growth is achieved by the local modification of the free energy of a GaAs~(001)~surface using spatially modulated strain fields caused by a 
partial oxidation of the underlying
 AlAs layer. A crucial parameter hereby is the aperture diameter which influences the strain distribution\cite{Strittmatter2012b}. At aperture diameters 
 close to 500~nm the subsequent growth of the strained InGaAs layer takes preferentially place at the local tensile strain maximum directly on top of the AlAs aperture located in the middle 
 of the mesas. Consequently, single QDs can be positioned in the middle of the mesas as is shown exemplary in the atomic force microscope image in Fig.~\ref{graph:0}~(b), where 2~QDs are positioned
  central to the aperture with a diameter of $\approx$~700~nm. This diameter also gives the upper bound of the alignment accuracy of SCQDs in the center of the larger mesa structures. 
Even though this alignment accuracy is lower than for SCQDs based on nanohole arrays, it is well suited for spatially deterministic integration of QDs into resonant-cavity LEDs~\cite{Unrau2012} and into micropillar cavities with a few $\mu\text{m}$ diameter and a comparatively large lateral mode profile. 
Statistical analysis shows that the probability of site-controlled nucleation, i.e. nucleation of a QD on top of the aperture as opposed to unwanted nucleation next to the aperture, is around 65\% measured over an area of 5\,$\mu \text{m}$\,x\,5\,$\mu \text{m}$ relevant for device processing of e.g. micropillar cavities.

\begin{figure}[htbp]
\includegraphics[width=0.95\linewidth]{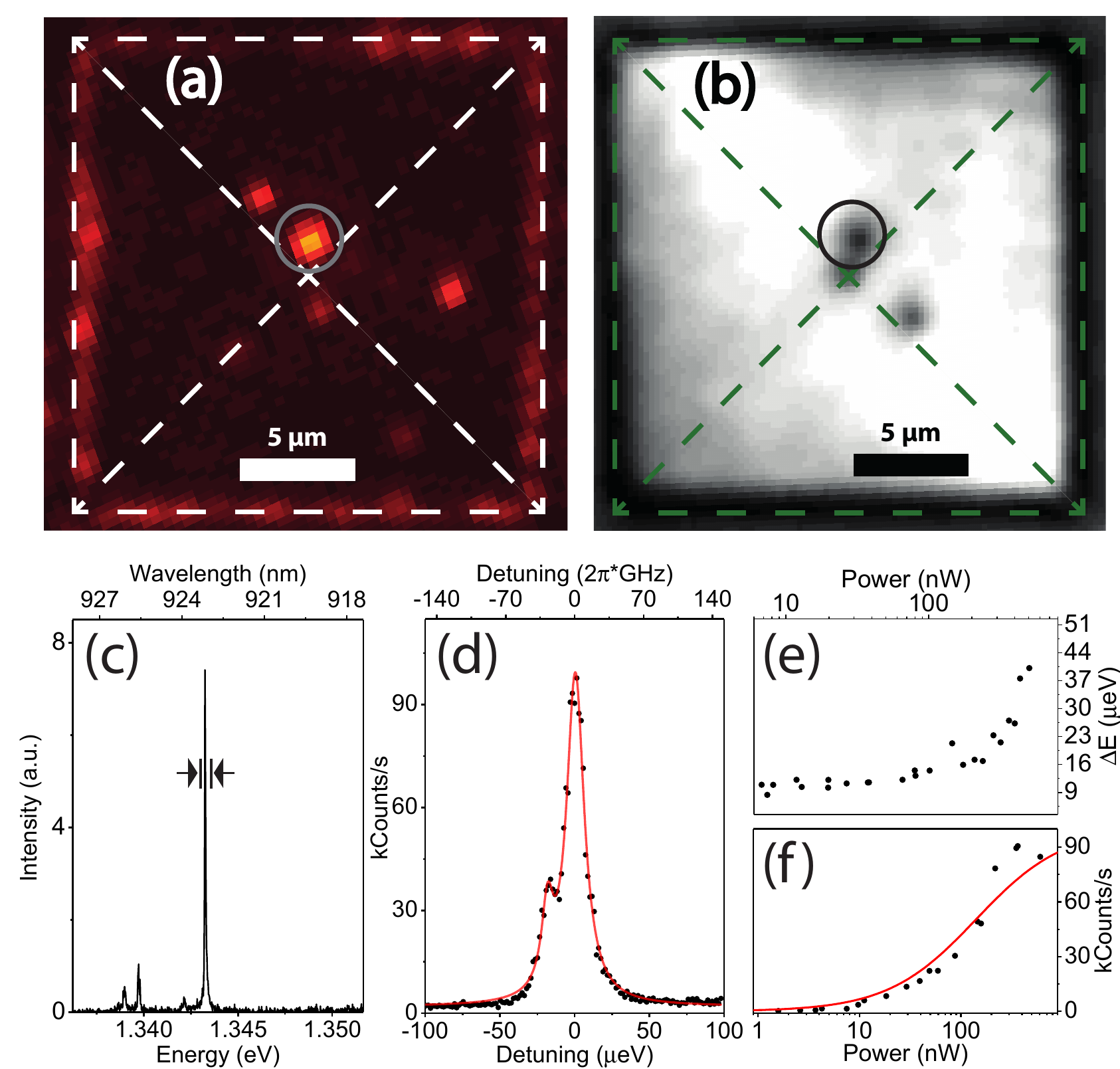}
\caption{\label{graph:1} (a) 2D $\mu$PL intensity map of a mesa containing a SCQD (marked by a circle) very close to its center.  (b) 2D cathodoluminescence intensity map of the same
mesa where the intensity is obtained by integrating over the emission from the
wetting layer (wavelength interval 900 to 915\,nm). The position of the SCQD under
study nicely correlates with the  area of
reduced wetting layer emission intensity, also indicated by a circle.
(c) Non-resonant $\mu$PL spectrum of the QD under investigation with an
excitonic (X) linewidth (FWHM) of 45\,$\mu$eV. (d) Resonance scan of the tunable laser across the X transition. 
(e) Linewidth dependence on the excitation power.
(f) Excitation power dependent resonance fluorescence intensity of the X transition. All spectra were taken at 5 K.}
\end{figure}

The sample was mounted inside a helium flow cryostat and cooled to 5 K. A linearly polarised beam of a tunable laser is focused onto the sample surface using a microscope objective (NA=0.65), which 
also collects the resultant fluorescence emitted by the QDs. Polarisation suppression of the laser with an extinction ratio of more than 6 orders of magnitude is achieved by using a
combination of a $\lambda /4-$plate and two polarising beam splitters. Ultimately, the fluorescence light is coupled into a single mode fiber for detection. Depending on the experimental requirements, 
the fiber is either directly connected to a single photon counting module (SPCM) or to a grating spectrometer with a spectral resolution of 25\,$\mu$eV. For reference measurements involving non-resonant 
excitation we use a laser diode emitting at 785\,nm. 

We pre-characterize the sample by spatially mapping the photoluminescence emitted
from QDs under non-resonant excitation (spectral range of plotted
intensity: 920 to 926\,nm). A typical $\mu$PL map scan of a mesa with an
oxide-aperture diameter of 700\,nm is presented in Fig.\,\ref{graph:1}. In addition to ensemble emission along the edges of the mesa with the size of 20\,$\mu\text{m}$ (side-length), we identify five
 distinct emission centers associated with QDs. The brightest emission originates from a SCQD close to the center of the mesa structure and aligned to the oxide aperture. Interestingly, the QD emission 
is spatially correlated to a local minimum of the wetting layer emission on top of the oxide aperture (cf. Fig.\,\ref{graph:1}\,(b)). In general, we use this type of mapping to visualize the (approximate) position 
of the oxide-aperture in the pre-characterization step. Fig.\,\ref{graph:1}\,(c)
shows a $\mu$PL spectrum of the target QD under non-resonant excitation at P =
1.4~\,$\mu$W. The dominating line is identified as neutral exciton  and has an
emission linewidth of 45\,$\mu$eV.
   
\begin{figure}[htbp]
\includegraphics[width=0.95\linewidth]{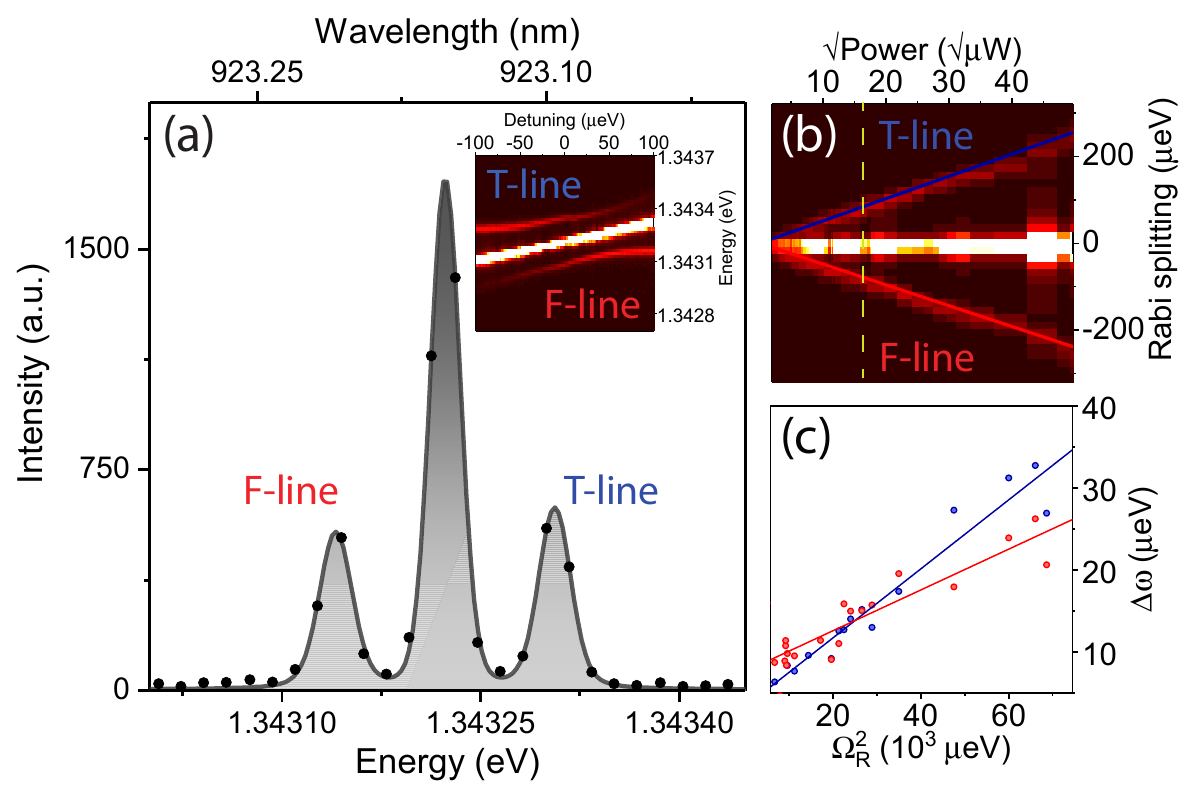}
\caption{\label{graph:2} (a) Emission spectrum of the target quantum dot under strong coherent excitation with P = 260\,$\mu$W (indicated by the dashed vertical (yellow) line in (b)). The characteristic Mollow-triplet evidences the coherent light-matter interaction 
under strong pump. Inset: Dispersive behaviour of the emission spectrum by tuning the laser across the resonance of the bare quantum dot. (b) Emission spectrum as a function of the square root of the excitation 
power. The Rabi splitting increases linearly with a slope of 5.1 (-4.8)\,$\mu \text{eV} /\sqrt{\mu \text{W}}$. (c) Blue (red) dots: linewidth (FWHM) of the T(F)- line as a function of excitation power.
A linear function $\Delta \omega \propto \chi \Omega_R^2 $is fit to the data
giving a slope of $\chi=$ 0.42 (0.26).}
\end{figure}  

Next, we sweep the tunable resonant laser across the transition indicated in Fig.\,\ref{graph:1}\,(c) and record the emitted fluorescence with a SPCM. A typical excitation spectrum 
at 240\,nW is shown in Fig.\,\ref{graph:1} (d). The linewidth for excitation powers below 10 nW is about 10\,$\mu$eV (cf. Fig.\,\ref{graph:1} (e))
which indicates non-Fourier limited photon emission. It is still one of the smallest values reported so far for SCQDs which supports the appealing prospects of the buried stressor approach. The emission linewidth could possibly be further improved by either using electrical fields to control the charge environment of the QD
\cite{Kuhlmann2015} or by reduction of the lifetime of the emitter via a cavity \cite{Wang2016}. The observed two lines correspond most likely to the
finestructure splitting of the exciton ($\Delta E_{\text{FSS}}=18.5$\,$\mu$eV). The finestructure splitting could be manipulated using piezo strain tuning. This technique has been used to suppress the fine for generating entangled photon pairs \cite{Trotta2014a} and is compatible with our planar sample geometry.
Increasing the power of the resonant laser we observe the typical saturation behaviour of a two-level
system (cf. Fig.\,\ref{graph:1}\,(f)). Fitting the data with
$I=\text{I}_{\infty}\text{P}/(\text{P}+\text{P}_{\text{sat}})$ where P is the
laser power, $\text{P}_{\text{sat}}$ the laser power at saturation and
$\text{I}_{\infty}$ the maximum fluorescence intensity, we obtain the laser
power at saturation of $\text{P}_{\text{sat}}$=\ 89\,nW.

In order to prove that we can coherently drive the target SCQD we perform excitation power dependent resonance fluorescence spectroscopy. The coherent interaction between the light field 
and the QD exciton is reflected in the appearance of a Mollow triplet in the emission spectrum under strong resonant excitation. At an excitation power of 260\,$\mu$W (cf. Fig.\,\ref{graph:2} (a)), the ratio between the sidepeak area and the 
central peak area is around 1:3, i.e. larger than the expected ratio of 1:2, which is probably caused by imperfect laser 
suppression as well as pure dephasing typical for solid state systems. We fit the sum of three independent Voigts to the spectrum fixing the linewith of the Gaussian contribution to the resolution 
limit of our spectrometer (25 $\mu eV$).  Observing the spectra as a function of the laser power incident on the quantum dot
 (see Fig.\,\ref{graph:2} (b)) reproduces the expected linear dependence of the F- and T-line on 
the amplitude of the electric field: 
the F- and T-line shift by $(-4.78\pm0.02)$\,$\mu\textrm{eV}/\sqrt{\mu \textrm{W}}$ and $(5.12 \pm  0.04)$\,$\mu\textrm{eV}/\sqrt{\mu \textrm{W}}$, respectively.
 The full width at half maximum (FWHM) of
 the sidepeaks is given by  \begin{align}\Delta
 \omega=\frac{3}{2}\Gamma_{1}+\gamma_{\text{PD}}+\gamma_{0}\label{eq:1} \end{align}
 where
 $\Gamma_{1}$ is the spontaneous radiative emission rate, $\gamma_{\text{PD}}$ is the
 phonon-induced pure dephasing rate and $\gamma_{0} $ is the non-phonon
 induced dephasing rate \cite{Wei2014}. Accordingly, the sidepeaks have been
 shown to be particularly susceptible to dephasing effects and can thus be used to
 identify and characterize the dephasing mechanisms
 \cite{Unsleber2015a,Wei2014,Ulrich2011}.
 The extracted linewidths of the sidepeaks as a function of the laser power are shown in Fig.\,\ref{graph:2}
 (c). 
 The linear increase observed in our experiment is consistent with pure
 dephasing induced by longitudinal acoustic (LA) phonons. Furthermore, using the y-axis intersection and the independently measured lifetime of our emitter $T_1=\frac{1}{\Gamma_1}= (898 \pm 41)$\,ps (measurement not shown), we obtain a non-phonon induced dephasing rate of $\gamma_0= (6.0 \pm 1.3)$\,$\mu$eV for vanishing pump power (i.e. $\gamma_{\text{PD}}=0$). 
 
\begin{figure}[htbp]
\includegraphics[width=0.95\linewidth]{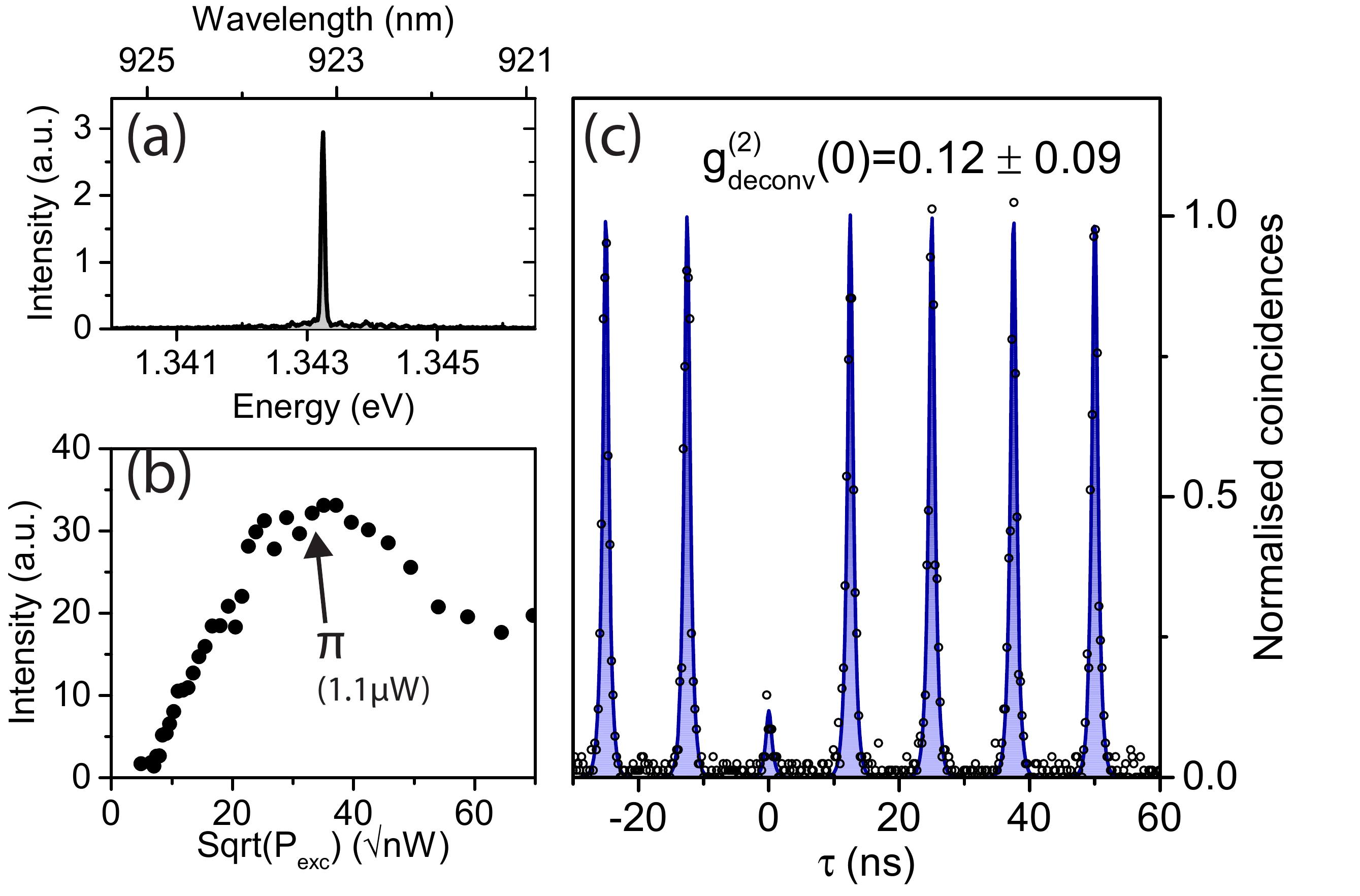}
\caption{\label{graph:3} (a) Resonance fluorescence emission spectrum of the quantum dot under $\pi$-pulse resonant excitation (1\,$\mu$W). (b) Intensity of the exciton emission as a function of 
the excitation power showing a Rabi-oscillation. The data point corresponding to the $\pi$-pulse area is indicated. (c) Photon auto-correlation measurement under $\pi$-pulse resonant excitation.}
\end{figure}

To demonstrate the quantum nature of emission, we perform photon statistics measurements of our site-controlled QDs under pulsed resonant excitation. For this purpose, we use a mode locked Ti:Sapphire laser providing pulses with a temporal width of 2\,ps (FWHM) at a repetition rate of 80\,MHz. A typical spectrum of the QD under pulsed excitation is 
shown in Fig.\,\ref{graph:3} (a) for an excitation power of 1\,$\mu$W. The single-line spectrum is very clean and the remaining background around the QD transition is attributed to 
imperfect laser suppression. Interestingly, under $\pi$-pulse excitation, the QD  exhibits an increased linewidth of about 45\,$\mu$eV. Measuring the intensity of the QD emission as a function of the laser power, we observe a Rabi oscillation up to about a 2$\pi$ pulse area (cf. Fig.\,\ref{graph:3} (b)) which 
is again a typical signature of the coherent control of the excitonic two-level system. The observed damping of the Rabi oscillation with increasing pulse area is probably caused by the coupling to 
longitudinal acoustic (LA) phonons \cite{Forstner2003,Ramsay2010}. 
  
The excitation power dependent RF measurements allow us to determine the $\pi$-pulse condition for which we study the photon autocorrelation function. The corresponding data is displayed in Fig.\,\ref{graph:3} (c) and the strong 
antibunching at zero time delay evidences single photon emission from a single
two-level system. A fit of the raw data with a convolution of an exponential
function (emitter decay) with the Gaussian instrument response function (IRF)
($\Delta t=350$\,ps) yields a $g^{(2)}(0)$ value of 0.12. We attribute to 
non-ideal value imperfect laser-suppression caused by the wavelength dependence of the used quarter waveplate.

In conclusion, we have shown that site-controlled QDs grown by the
buried-stressor growth technique exhibit promising optical properties under resonant excitation.
We evidence the coherent interaction between exciton and laser field by
observing the Mollow triplet sustained up to Rabi splittings of ~200 $\mu$eV.
Using resonant pulsed excitation, we prove single-photon emission by performing
an intensity autocorrelation measurement obtaining $g^{(2)}(0)$ = 0.12. Our
results demonstrates the high potential of this approach for the realization of scalable quantum devices based on high-quality single photon emitters. Further development could aim at dense arrays of site-controlled buried-stressor QDs and at their integration in microlenses for enhanced photon-extraction efficiency. The
 buried-stressor approach is also very suitable for the realization of micropillar lasers with a controlled number of QDs at the anti-node of the fundamental cavity mode. 

The research leading to these results has received funding from the German Research Foundation via CRC 787 and Re2974/5-1, the Volkswagen Foundation via the project NeuroQNet and from the European Research Council under the European Union's Seventh Framework ERC Grant Agreement No. 615613. 


\begin{thebibliography}{27}%
\makeatletter
\providecommand \@ifxundefined [1]{%
 \@ifx{#1\undefined}
}%
\providecommand \@ifnum [1]{%
 \ifnum #1\expandafter \@firstoftwo
 \else \expandafter \@secondoftwo
 \fi
}%
\providecommand \@ifx [1]{%
 \ifx #1\expandafter \@firstoftwo
 \else \expandafter \@secondoftwo
 \fi
}%
\providecommand \natexlab [1]{#1}%
\providecommand \enquote  [1]{``#1''}%
\providecommand \bibnamefont  [1]{#1}%
\providecommand \bibfnamefont [1]{#1}%
\providecommand \citenamefont [1]{#1}%
\providecommand \href@noop [0]{\@secondoftwo}%
\providecommand \href [0]{\begingroup \@sanitize@url \@href}%
\providecommand \@href[1]{\@@startlink{#1}\@@href}%
\providecommand \@@href[1]{\endgroup#1\@@endlink}%
\providecommand \@sanitize@url [0]{\catcode `\\12\catcode `\$12\catcode
  `\&12\catcode `\#12\catcode `\^12\catcode `\_12\catcode `\%12\relax}%
\providecommand \@@startlink[1]{}%
\providecommand \@@endlink[0]{}%
\providecommand \url  [0]{\begingroup\@sanitize@url \@url }%
\providecommand \@url [1]{\endgroup\@href {#1}{\urlprefix }}%
\providecommand \urlprefix  [0]{URL }%
\providecommand \Eprint [0]{\href }%
\providecommand \doibase [0]{http://dx.doi.org/}%
\providecommand \selectlanguage [0]{\@gobble}%
\providecommand \bibinfo  [0]{\@secondoftwo}%
\providecommand \bibfield  [0]{\@secondoftwo}%
\providecommand \translation [1]{[#1]}%
\providecommand \BibitemOpen [0]{}%
\providecommand \bibitemStop [0]{}%
\providecommand \bibitemNoStop [0]{.\EOS\space}%
\providecommand \EOS [0]{\spacefactor3000\relax}%
\providecommand \BibitemShut  [1]{\csname bibitem#1\endcsname}%
\let\auto@bib@innerbib\@empty
\bibitem [{\citenamefont {Kimble}(2008)}]{Kimble2008}%
  \BibitemOpen
  \bibfield  {author} {\bibinfo {author} {\bibfnamefont {H.~J.}\ \bibnamefont
  {Kimble}},\ }\href {\doibase 10.1038/nature07127} {\bibfield  {journal}
  {\bibinfo  {journal} {Nature}\ }\textbf {\bibinfo {volume} {453}},\ \bibinfo
  {pages} {1023} (\bibinfo {year} {2008})}\BibitemShut {NoStop}%
\bibitem [{\citenamefont {Gisin}\ and\ \citenamefont {Thew}(2007)}]{Gisin2007}%
  \BibitemOpen
  \bibfield  {author} {\bibinfo {author} {\bibfnamefont {N.}~\bibnamefont
  {Gisin}}\ and\ \bibinfo {author} {\bibfnamefont {R.}~\bibnamefont {Thew}},\
  }\href {\doibase 10.1038/nphoton.2007.22} {\bibfield  {journal} {\bibinfo
  {journal} {Nat Photon}\ }\textbf {\bibinfo {volume} {1}},\ \bibinfo {pages}
  {165} (\bibinfo {year} {2007})}\BibitemShut {NoStop}%
\bibitem [{\citenamefont {Sangouard}\ \emph {et~al.}(2007)\citenamefont
  {Sangouard}, \citenamefont {Simon}, \citenamefont {Min{\'{a}}ř},
  \citenamefont {Zbinden}, \citenamefont {{De Riedmatten}},\ and\ \citenamefont
  {Gisin}}]{Sangouard2007}%
  \BibitemOpen
  \bibfield  {author} {\bibinfo {author} {\bibfnamefont {N.}~\bibnamefont
  {Sangouard}}, \bibinfo {author} {\bibfnamefont {C.}~\bibnamefont {Simon}},
  \bibinfo {author} {\bibfnamefont {J.}~\bibnamefont {Min{\'{a}}ř}}, \bibinfo
  {author} {\bibfnamefont {H.}~\bibnamefont {Zbinden}}, \bibinfo {author}
  {\bibfnamefont {H.}~\bibnamefont {{De Riedmatten}}}, \ and\ \bibinfo {author}
  {\bibfnamefont {N.}~\bibnamefont {Gisin}},\ }\href {\doibase
  10.1103/PhysRevA.76.050301} {\bibfield  {journal} {\bibinfo  {journal}
  {Physical Review A - Atomic, Molecular, and Optical Physics}\ }\textbf
  {\bibinfo {volume} {76}},\ \bibinfo {pages} {1} (\bibinfo {year}
  {2007})}\BibitemShut {NoStop}%
\bibitem [{\citenamefont {Ding}\ \emph {et~al.}(2016)\citenamefont {Ding},
  \citenamefont {He}, \citenamefont {Duan}, \citenamefont {Gregersen},
  \citenamefont {Chen}, \citenamefont {Unsleber}, \citenamefont {Maier},
  \citenamefont {Schneider}, \citenamefont {Kamp}, \citenamefont
  {H{\"{o}}fling}, \citenamefont {Lu},\ and\ \citenamefont {Pan}}]{Ding2016}%
  \BibitemOpen
  \bibfield  {author} {\bibinfo {author} {\bibfnamefont {X.}~\bibnamefont
  {Ding}}, \bibinfo {author} {\bibfnamefont {Y.}~\bibnamefont {He}}, \bibinfo
  {author} {\bibfnamefont {Z.~C.}\ \bibnamefont {Duan}}, \bibinfo {author}
  {\bibfnamefont {N.}~\bibnamefont {Gregersen}}, \bibinfo {author}
  {\bibfnamefont {M.~C.}\ \bibnamefont {Chen}}, \bibinfo {author}
  {\bibfnamefont {S.}~\bibnamefont {Unsleber}}, \bibinfo {author}
  {\bibfnamefont {S.}~\bibnamefont {Maier}}, \bibinfo {author} {\bibfnamefont
  {C.}~\bibnamefont {Schneider}}, \bibinfo {author} {\bibfnamefont
  {M.}~\bibnamefont {Kamp}}, \bibinfo {author} {\bibfnamefont {S.}~\bibnamefont
  {H{\"{o}}fling}}, \bibinfo {author} {\bibfnamefont {C.~Y.}\ \bibnamefont
  {Lu}}, \ and\ \bibinfo {author} {\bibfnamefont {J.~W.}\ \bibnamefont {Pan}},\
  }\href {\doibase 10.1103/PhysRevLett.116.020401} {\bibfield  {journal}
  {\bibinfo  {journal} {Physical Review Letters}\ }\textbf {\bibinfo {volume}
  {116}},\ \bibinfo {pages} {1} (\bibinfo {year} {2016})}\BibitemShut {NoStop}%
\bibitem [{\citenamefont {Kwiat}\ \emph {et~al.}(1995)\citenamefont {Kwiat},
  \citenamefont {Mattle}, \citenamefont {Weinfurter}, \citenamefont
  {Zeilinger}, \citenamefont {Sergienko},\ and\ \citenamefont
  {Shih}}]{Kwiat1995}%
  \BibitemOpen
  \bibfield  {author} {\bibinfo {author} {\bibfnamefont {P.~G.}\ \bibnamefont
  {Kwiat}}, \bibinfo {author} {\bibfnamefont {K.}~\bibnamefont {Mattle}},
  \bibinfo {author} {\bibfnamefont {H.}~\bibnamefont {Weinfurter}}, \bibinfo
  {author} {\bibfnamefont {A.}~\bibnamefont {Zeilinger}}, \bibinfo {author}
  {\bibfnamefont {A.~V.}\ \bibnamefont {Sergienko}}, \ and\ \bibinfo {author}
  {\bibfnamefont {Y.}~\bibnamefont {Shih}},\ }\href {\doibase
  10.1103/PhysRevLett.75.4337} {\bibfield  {journal} {\bibinfo  {journal}
  {Physical Review Letters}\ }\textbf {\bibinfo {volume} {75}},\ \bibinfo
  {pages} {4337} (\bibinfo {year} {1995})}\BibitemShut {NoStop}%
\bibitem [{\citenamefont {Dousse}\ \emph {et~al.}(2008)\citenamefont {Dousse},
  \citenamefont {Lanco}, \citenamefont {Suffczynski}, \citenamefont {Semenova},
  \citenamefont {Miard}, \citenamefont {Lema{\^{i}}tre}, \citenamefont
  {Sagnes}, \citenamefont {Roblin}, \citenamefont {Bloch},\ and\ \citenamefont
  {Senellart}}]{Dousse2008}%
  \BibitemOpen
  \bibfield  {author} {\bibinfo {author} {\bibfnamefont {A.}~\bibnamefont
  {Dousse}}, \bibinfo {author} {\bibfnamefont {L.}~\bibnamefont {Lanco}},
  \bibinfo {author} {\bibfnamefont {J.}~\bibnamefont {Suffczynski}}, \bibinfo
  {author} {\bibfnamefont {E.}~\bibnamefont {Semenova}}, \bibinfo {author}
  {\bibfnamefont {A.}~\bibnamefont {Miard}}, \bibinfo {author} {\bibfnamefont
  {A.}~\bibnamefont {Lema{\^{i}}tre}}, \bibinfo {author} {\bibfnamefont
  {I.}~\bibnamefont {Sagnes}}, \bibinfo {author} {\bibfnamefont
  {C.}~\bibnamefont {Roblin}}, \bibinfo {author} {\bibfnamefont
  {J.}~\bibnamefont {Bloch}}, \ and\ \bibinfo {author} {\bibfnamefont
  {P.}~\bibnamefont {Senellart}},\ }\href {\doibase
  10.1103/PhysRevLett.101.267404} {\bibfield  {journal} {\bibinfo  {journal}
  {Physical Review Letters}\ }\textbf {\bibinfo {volume} {101}},\ \bibinfo
  {pages} {30} (\bibinfo {year} {2008})}\BibitemShut {NoStop}%
\bibitem [{\citenamefont {Gschrey}\ \emph {et~al.}(2013)\citenamefont
  {Gschrey}, \citenamefont {Gericke}, \citenamefont {Sch{\"{u}}{\ss}ler},
  \citenamefont {Schmidt}, \citenamefont {Schulze}, \citenamefont {Heindel},
  \citenamefont {Rodt}, \citenamefont {Strittmatter},\ and\ \citenamefont
  {Reitzenstein}}]{Gschrey2013}%
  \BibitemOpen
  \bibfield  {author} {\bibinfo {author} {\bibfnamefont {M.}~\bibnamefont
  {Gschrey}}, \bibinfo {author} {\bibfnamefont {F.}~\bibnamefont {Gericke}},
  \bibinfo {author} {\bibfnamefont {A.}~\bibnamefont {Sch{\"{u}}{\ss}ler}},
  \bibinfo {author} {\bibfnamefont {R.}~\bibnamefont {Schmidt}}, \bibinfo
  {author} {\bibfnamefont {J.~H.}\ \bibnamefont {Schulze}}, \bibinfo {author}
  {\bibfnamefont {T.}~\bibnamefont {Heindel}}, \bibinfo {author} {\bibfnamefont
  {S.}~\bibnamefont {Rodt}}, \bibinfo {author} {\bibfnamefont {A.}~\bibnamefont
  {Strittmatter}}, \ and\ \bibinfo {author} {\bibfnamefont {S.}~\bibnamefont
  {Reitzenstein}},\ }\href@noop {} {\bibfield  {journal} {\bibinfo  {journal}
  {Applied Physics Letters}\ }\textbf {\bibinfo {volume} {102}} (\bibinfo
  {year} {2013})}\BibitemShut {NoStop}%
\bibitem [{\citenamefont {Laucht}\ \emph {et~al.}(2012)\citenamefont {Laucht},
  \citenamefont {P{\"{u}}tz}, \citenamefont {G{\"{u}}nthner}, \citenamefont
  {Hauke}, \citenamefont {Saive}, \citenamefont {Fr{\'{e}}d{\'{e}}rick},
  \citenamefont {Bichler}, \citenamefont {Amann}, \citenamefont {Holleitner},
  \citenamefont {Kaniber},\ and\ \citenamefont {Finley}}]{Laucht2012}%
  \BibitemOpen
  \bibfield  {author} {\bibinfo {author} {\bibfnamefont {A.}~\bibnamefont
  {Laucht}}, \bibinfo {author} {\bibfnamefont {S.}~\bibnamefont {P{\"{u}}tz}},
  \bibinfo {author} {\bibfnamefont {T.}~\bibnamefont {G{\"{u}}nthner}},
  \bibinfo {author} {\bibfnamefont {N.}~\bibnamefont {Hauke}}, \bibinfo
  {author} {\bibfnamefont {R.}~\bibnamefont {Saive}}, \bibinfo {author}
  {\bibfnamefont {S.}~\bibnamefont {Fr{\'{e}}d{\'{e}}rick}}, \bibinfo {author}
  {\bibfnamefont {M.}~\bibnamefont {Bichler}}, \bibinfo {author} {\bibfnamefont
  {M.-C.}\ \bibnamefont {Amann}}, \bibinfo {author} {\bibfnamefont {A.~W.}\
  \bibnamefont {Holleitner}}, \bibinfo {author} {\bibfnamefont
  {M.}~\bibnamefont {Kaniber}}, \ and\ \bibinfo {author} {\bibfnamefont
  {J.~J.}\ \bibnamefont {Finley}},\ }\href {\doibase 10.1103/PhysRevX.2.011014}
  {\bibfield  {journal} {\bibinfo  {journal} {Physical Review X}\ }\textbf
  {\bibinfo {volume} {2}},\ \bibinfo {pages} {011014} (\bibinfo {year}
  {2012})}\BibitemShut {NoStop}%
\bibitem [{\citenamefont {Makhonin}\ \emph {et~al.}(2014)\citenamefont
  {Makhonin}, \citenamefont {Dixon}, \citenamefont {Coles}, \citenamefont
  {Royall}, \citenamefont {Luxmoore}, \citenamefont {Clarke}, \citenamefont
  {Hugues}, \citenamefont {Skolnick},\ and\ \citenamefont
  {Fox}}]{Makhonin2014}%
  \BibitemOpen
  \bibfield  {author} {\bibinfo {author} {\bibfnamefont {M.~N.}\ \bibnamefont
  {Makhonin}}, \bibinfo {author} {\bibfnamefont {J.~E.}\ \bibnamefont {Dixon}},
  \bibinfo {author} {\bibfnamefont {R.~J.}\ \bibnamefont {Coles}}, \bibinfo
  {author} {\bibfnamefont {B.}~\bibnamefont {Royall}}, \bibinfo {author}
  {\bibfnamefont {I.~J.}\ \bibnamefont {Luxmoore}}, \bibinfo {author}
  {\bibfnamefont {E.}~\bibnamefont {Clarke}}, \bibinfo {author} {\bibfnamefont
  {M.}~\bibnamefont {Hugues}}, \bibinfo {author} {\bibfnamefont {M.~S.}\
  \bibnamefont {Skolnick}}, \ and\ \bibinfo {author} {\bibfnamefont {A.~M.}\
  \bibnamefont {Fox}},\ }\href {\doibase 10.1021/nl5032937} {\bibfield
  {journal} {\bibinfo  {journal} {Nano Letters}\ }\textbf {\bibinfo {volume}
  {14}},\ \bibinfo {pages} {6997} (\bibinfo {year} {2014})}\BibitemShut
  {NoStop}%
\bibitem [{\citenamefont {Pelucchi}\ \emph {et~al.}(2007)\citenamefont
  {Pelucchi}, \citenamefont {Watanabe}, \citenamefont {Leifer}, \citenamefont
  {Zhu}, \citenamefont {Dwir}, \citenamefont {{De Los Rios}},\ and\
  \citenamefont {Kapon}}]{Pelucchi2007}%
  \BibitemOpen
  \bibfield  {author} {\bibinfo {author} {\bibfnamefont {E.}~\bibnamefont
  {Pelucchi}}, \bibinfo {author} {\bibfnamefont {S.}~\bibnamefont {Watanabe}},
  \bibinfo {author} {\bibfnamefont {K.}~\bibnamefont {Leifer}}, \bibinfo
  {author} {\bibfnamefont {Q.}~\bibnamefont {Zhu}}, \bibinfo {author}
  {\bibfnamefont {B.}~\bibnamefont {Dwir}}, \bibinfo {author} {\bibfnamefont
  {P.}~\bibnamefont {{De Los Rios}}}, \ and\ \bibinfo {author} {\bibfnamefont
  {E.}~\bibnamefont {Kapon}},\ }\href {\doibase 10.1021/nl0702012} {\bibfield
  {journal} {\bibinfo  {journal} {Nano Letters}\ }\textbf {\bibinfo {volume}
  {7}},\ \bibinfo {pages} {1282} (\bibinfo {year} {2007})}\BibitemShut
  {NoStop}%
\bibitem [{\citenamefont {Surrente}\ \emph {et~al.}(2009)\citenamefont
  {Surrente}, \citenamefont {Gallo}, \citenamefont {Felici}, \citenamefont
  {Dwir}, \citenamefont {Rudra},\ and\ \citenamefont {Kapon}}]{Surrente2009}%
  \BibitemOpen
  \bibfield  {author} {\bibinfo {author} {\bibfnamefont {A.}~\bibnamefont
  {Surrente}}, \bibinfo {author} {\bibfnamefont {P.}~\bibnamefont {Gallo}},
  \bibinfo {author} {\bibfnamefont {M.}~\bibnamefont {Felici}}, \bibinfo
  {author} {\bibfnamefont {B.}~\bibnamefont {Dwir}}, \bibinfo {author}
  {\bibfnamefont {A.}~\bibnamefont {Rudra}}, \ and\ \bibinfo {author}
  {\bibfnamefont {E.}~\bibnamefont {Kapon}},\ }\href {\doibase
  10.1088/0957-4484/20/41/415205} {\bibfield  {journal} {\bibinfo  {journal}
  {Nanotechnology}\ }\textbf {\bibinfo {volume} {20}},\ \bibinfo {pages}
  {415205} (\bibinfo {year} {2009})}\BibitemShut {NoStop}%
\bibitem [{\citenamefont {Schneider}\ \emph {et~al.}(2009)\citenamefont
  {Schneider}, \citenamefont {Heindel}, \citenamefont {Huggenberger},
  \citenamefont {Weinmann}, \citenamefont {Kistner}, \citenamefont {Kamp},
  \citenamefont {Reitzenstein}, \citenamefont {H{\"{o}}fling},\ and\
  \citenamefont {Forchel}}]{Schneider2009a}%
  \BibitemOpen
  \bibfield  {author} {\bibinfo {author} {\bibfnamefont {C.}~\bibnamefont
  {Schneider}}, \bibinfo {author} {\bibfnamefont {T.}~\bibnamefont {Heindel}},
  \bibinfo {author} {\bibfnamefont {A.}~\bibnamefont {Huggenberger}}, \bibinfo
  {author} {\bibfnamefont {P.}~\bibnamefont {Weinmann}}, \bibinfo {author}
  {\bibfnamefont {C.}~\bibnamefont {Kistner}}, \bibinfo {author} {\bibfnamefont
  {M.}~\bibnamefont {Kamp}}, \bibinfo {author} {\bibfnamefont {S.}~\bibnamefont
  {Reitzenstein}}, \bibinfo {author} {\bibfnamefont {S.}~\bibnamefont
  {H{\"{o}}fling}}, \ and\ \bibinfo {author} {\bibfnamefont {A.}~\bibnamefont
  {Forchel}},\ }\href {\doibase 10.1063/1.3097016} {\bibfield  {journal}
  {\bibinfo  {journal} {Applied Physics Letters}\ }\textbf {\bibinfo {volume}
  {94}},\ \bibinfo {pages} {3} (\bibinfo {year} {2009})}\BibitemShut {NoStop}%
\bibitem [{\citenamefont {Pfau}\ \emph {et~al.}(2009)\citenamefont {Pfau},
  \citenamefont {Gushterov}, \citenamefont {Reithmaier}, \citenamefont
  {Cestier}, \citenamefont {Eisenstein}, \citenamefont {Linder},\ and\
  \citenamefont {Gershoni}}]{Pfau2009}%
  \BibitemOpen
  \bibfield  {author} {\bibinfo {author} {\bibfnamefont {T.~J.}\ \bibnamefont
  {Pfau}}, \bibinfo {author} {\bibfnamefont {A.}~\bibnamefont {Gushterov}},
  \bibinfo {author} {\bibfnamefont {J.~P.}\ \bibnamefont {Reithmaier}},
  \bibinfo {author} {\bibfnamefont {I.}~\bibnamefont {Cestier}}, \bibinfo
  {author} {\bibfnamefont {G.}~\bibnamefont {Eisenstein}}, \bibinfo {author}
  {\bibfnamefont {E.}~\bibnamefont {Linder}}, \ and\ \bibinfo {author}
  {\bibfnamefont {D.}~\bibnamefont {Gershoni}},\ }\href {\doibase
  10.1063/1.3265918} {\bibfield  {journal} {\bibinfo  {journal} {Applied
  Physics Letters}\ }\textbf {\bibinfo {volume} {95}},\ \bibinfo {pages} {93}
  (\bibinfo {year} {2009})}\BibitemShut {NoStop}%
\bibitem [{\citenamefont {Albert}\ \emph {et~al.}(2010)\citenamefont {Albert},
  \citenamefont {Stobbe}, \citenamefont {Schneider}, \citenamefont {Heindel},
  \citenamefont {Reitzenstein}, \citenamefont {H{\"{o}}fling}, \citenamefont
  {Lodahl}, \citenamefont {Worschech},\ and\ \citenamefont
  {Forchel}}]{Albert2010}%
  \BibitemOpen
  \bibfield  {author} {\bibinfo {author} {\bibfnamefont {F.}~\bibnamefont
  {Albert}}, \bibinfo {author} {\bibfnamefont {S.}~\bibnamefont {Stobbe}},
  \bibinfo {author} {\bibfnamefont {C.}~\bibnamefont {Schneider}}, \bibinfo
  {author} {\bibfnamefont {T.}~\bibnamefont {Heindel}}, \bibinfo {author}
  {\bibfnamefont {S.}~\bibnamefont {Reitzenstein}}, \bibinfo {author}
  {\bibfnamefont {S.}~\bibnamefont {H{\"{o}}fling}}, \bibinfo {author}
  {\bibfnamefont {P.}~\bibnamefont {Lodahl}}, \bibinfo {author} {\bibfnamefont
  {L.}~\bibnamefont {Worschech}}, \ and\ \bibinfo {author} {\bibfnamefont
  {A.}~\bibnamefont {Forchel}},\ }\href@noop {} {\bibfield  {journal} {\bibinfo
   {journal} {Applied Physics Letters}\ }\textbf {\bibinfo {volume} {96}}
  (\bibinfo {year} {2010})}\BibitemShut {NoStop}%
\bibitem [{\citenamefont {J{\"{o}}ns}\ \emph {et~al.}(2013)\citenamefont
  {J{\"{o}}ns}, \citenamefont {Atkinson}, \citenamefont {M{\"{u}}ller},
  \citenamefont {Heldmaier}, \citenamefont {Ulrich}, \citenamefont {Schmidt},\
  and\ \citenamefont {Michler}}]{Jons2013}%
  \BibitemOpen
  \bibfield  {author} {\bibinfo {author} {\bibfnamefont {K.~D.}\ \bibnamefont
  {J{\"{o}}ns}}, \bibinfo {author} {\bibfnamefont {P.}~\bibnamefont
  {Atkinson}}, \bibinfo {author} {\bibfnamefont {M.}~\bibnamefont
  {M{\"{u}}ller}}, \bibinfo {author} {\bibfnamefont {M.}~\bibnamefont
  {Heldmaier}}, \bibinfo {author} {\bibfnamefont {S.~M.}\ \bibnamefont
  {Ulrich}}, \bibinfo {author} {\bibfnamefont {O.~G.}\ \bibnamefont {Schmidt}},
  \ and\ \bibinfo {author} {\bibfnamefont {P.}~\bibnamefont {Michler}},\ }\href
  {\doibase 10.1021/nl303668z} {\bibfield  {journal} {\bibinfo  {journal} {Nano
  Letters}\ }\textbf {\bibinfo {volume} {13}},\ \bibinfo {pages} {126}
  (\bibinfo {year} {2013})}\BibitemShut {NoStop}%
\bibitem [{\citenamefont {Unsleber}\ \emph {et~al.}(2015)\citenamefont
  {Unsleber}, \citenamefont {Maier}, \citenamefont {McCutcheon}, \citenamefont
  {He}, \citenamefont {Dambach}, \citenamefont {Gschrey}, \citenamefont
  {Gregersen}, \citenamefont {M{\o}rk}, \citenamefont {Reitzenstein},
  \citenamefont {H{\"{o}}fling}, \citenamefont {Schneider},\ and\ \citenamefont
  {Kamp}}]{Unsleber2015a}%
  \BibitemOpen
  \bibfield  {author} {\bibinfo {author} {\bibfnamefont {S.}~\bibnamefont
  {Unsleber}}, \bibinfo {author} {\bibfnamefont {S.}~\bibnamefont {Maier}},
  \bibinfo {author} {\bibfnamefont {D.~P.~S.}\ \bibnamefont {McCutcheon}},
  \bibinfo {author} {\bibfnamefont {Y.-M.}\ \bibnamefont {He}}, \bibinfo
  {author} {\bibfnamefont {M.}~\bibnamefont {Dambach}}, \bibinfo {author}
  {\bibfnamefont {M.}~\bibnamefont {Gschrey}}, \bibinfo {author} {\bibfnamefont
  {N.}~\bibnamefont {Gregersen}}, \bibinfo {author} {\bibfnamefont
  {J.}~\bibnamefont {M{\o}rk}}, \bibinfo {author} {\bibfnamefont
  {S.}~\bibnamefont {Reitzenstein}}, \bibinfo {author} {\bibfnamefont
  {S.}~\bibnamefont {H{\"{o}}fling}}, \bibinfo {author} {\bibfnamefont
  {C.}~\bibnamefont {Schneider}}, \ and\ \bibinfo {author} {\bibfnamefont
  {M.}~\bibnamefont {Kamp}},\ }\href {\doibase 10.1364/OPTICA.2.001072}
  {\bibfield  {journal} {\bibinfo  {journal} {Optica}\ }\textbf {\bibinfo
  {volume} {2}},\ \bibinfo {pages} {1072} (\bibinfo {year} {2015})}\BibitemShut
  {NoStop}%
\bibitem [{\citenamefont {Huber}\ \emph {et~al.}(2014)\citenamefont {Huber},
  \citenamefont {Predojevic}, \citenamefont {Khoshnegar}, \citenamefont
  {Dalacu}, \citenamefont {Poole}, \citenamefont {Majedi},\ and\ \citenamefont
  {Weihs}}]{Huber2014}%
  \BibitemOpen
  \bibfield  {author} {\bibinfo {author} {\bibfnamefont {T.}~\bibnamefont
  {Huber}}, \bibinfo {author} {\bibfnamefont {A.}~\bibnamefont {Predojevic}},
  \bibinfo {author} {\bibfnamefont {M.}~\bibnamefont {Khoshnegar}}, \bibinfo
  {author} {\bibfnamefont {D.}~\bibnamefont {Dalacu}}, \bibinfo {author}
  {\bibfnamefont {P.~J.}\ \bibnamefont {Poole}}, \bibinfo {author}
  {\bibfnamefont {H.}~\bibnamefont {Majedi}}, \ and\ \bibinfo {author}
  {\bibfnamefont {G.}~\bibnamefont {Weihs}},\ }\href {\doibase
  10.1021/nl503581d} {\bibfield  {journal} {\bibinfo  {journal} {Nano Letters}\
  }\textbf {\bibinfo {volume} {14}},\ \bibinfo {pages} {7107} (\bibinfo {year}
  {2014})}\BibitemShut {NoStop}%
\bibitem [{\citenamefont {Unrau}\ \emph {et~al.}(2012)\citenamefont {Unrau},
  \citenamefont {Quandt}, \citenamefont {Schulze}, \citenamefont {Heindel},
  \citenamefont {Germann}, \citenamefont {Hitzemann}, \citenamefont
  {Strittmatter}, \citenamefont {Reitzenstein}, \citenamefont {Pohl},\ and\
  \citenamefont {Bimberg}}]{Unrau2012}%
  \BibitemOpen
  \bibfield  {author} {\bibinfo {author} {\bibfnamefont {W.}~\bibnamefont
  {Unrau}}, \bibinfo {author} {\bibfnamefont {D.}~\bibnamefont {Quandt}},
  \bibinfo {author} {\bibfnamefont {J.~H.}\ \bibnamefont {Schulze}}, \bibinfo
  {author} {\bibfnamefont {T.}~\bibnamefont {Heindel}}, \bibinfo {author}
  {\bibfnamefont {T.~D.}\ \bibnamefont {Germann}}, \bibinfo {author}
  {\bibfnamefont {O.}~\bibnamefont {Hitzemann}}, \bibinfo {author}
  {\bibfnamefont {A.}~\bibnamefont {Strittmatter}}, \bibinfo {author}
  {\bibfnamefont {S.}~\bibnamefont {Reitzenstein}}, \bibinfo {author}
  {\bibfnamefont {U.~W.}\ \bibnamefont {Pohl}}, \ and\ \bibinfo {author}
  {\bibfnamefont {D.}~\bibnamefont {Bimberg}},\ }\href {\doibase
  10.1063/1.4767525} {\bibfield  {journal} {\bibinfo  {journal} {Applied
  Physics Letters}\ }\textbf {\bibinfo {volume} {101}},\ \bibinfo {pages}
  {211119} (\bibinfo {year} {2012})}\BibitemShut {NoStop}%
\bibitem [{\citenamefont {Strittmatter}\ \emph
  {et~al.}(2012{\natexlab{a}})\citenamefont {Strittmatter}, \citenamefont
  {Holzbecher}, \citenamefont {Schliwa}, \citenamefont {Schulze}, \citenamefont
  {Quandt}, \citenamefont {Germann}, \citenamefont {Dreismann}, \citenamefont
  {Hitzemann}, \citenamefont {Stock}, \citenamefont {Ostapenko}, \citenamefont
  {Rodt}, \citenamefont {Unrau}, \citenamefont {Pohl}, \citenamefont
  {Hoffmann}, \citenamefont {Bimberg},\ and\ \citenamefont
  {Haisler}}]{Strittmatter2012b}%
  \BibitemOpen
  \bibfield  {author} {\bibinfo {author} {\bibfnamefont {A.}~\bibnamefont
  {Strittmatter}}, \bibinfo {author} {\bibfnamefont {A.}~\bibnamefont
  {Holzbecher}}, \bibinfo {author} {\bibfnamefont {A.}~\bibnamefont {Schliwa}},
  \bibinfo {author} {\bibfnamefont {J.~H.}\ \bibnamefont {Schulze}}, \bibinfo
  {author} {\bibfnamefont {D.}~\bibnamefont {Quandt}}, \bibinfo {author}
  {\bibfnamefont {T.~D.}\ \bibnamefont {Germann}}, \bibinfo {author}
  {\bibfnamefont {A.}~\bibnamefont {Dreismann}}, \bibinfo {author}
  {\bibfnamefont {O.}~\bibnamefont {Hitzemann}}, \bibinfo {author}
  {\bibfnamefont {E.}~\bibnamefont {Stock}}, \bibinfo {author} {\bibfnamefont
  {I.~A.}\ \bibnamefont {Ostapenko}}, \bibinfo {author} {\bibfnamefont
  {S.}~\bibnamefont {Rodt}}, \bibinfo {author} {\bibfnamefont {W.}~\bibnamefont
  {Unrau}}, \bibinfo {author} {\bibfnamefont {U.~W.}\ \bibnamefont {Pohl}},
  \bibinfo {author} {\bibfnamefont {A.}~\bibnamefont {Hoffmann}}, \bibinfo
  {author} {\bibfnamefont {D.}~\bibnamefont {Bimberg}}, \ and\ \bibinfo
  {author} {\bibfnamefont {V.}~\bibnamefont {Haisler}},\ }\href {\doibase
  10.1002/pssa.201228407} {\bibfield  {journal} {\bibinfo  {journal} {Physica
  Status Solidi (A) Applications and Materials Science}\ }\textbf {\bibinfo
  {volume} {209}},\ \bibinfo {pages} {2411} (\bibinfo {year}
  {2012}{\natexlab{a}})}\BibitemShut {NoStop}%
\bibitem [{\citenamefont {Strittmatter}\ \emph
  {et~al.}(2012{\natexlab{b}})\citenamefont {Strittmatter}, \citenamefont
  {Schliwa}, \citenamefont {Schulze}, \citenamefont {Germann}, \citenamefont
  {Dreismann}, \citenamefont {Hitzemann}, \citenamefont {Stock}, \citenamefont
  {Ostapenko}, \citenamefont {Rodt}, \citenamefont {Unrau}, \citenamefont
  {Pohl}, \citenamefont {Hoffmann}, \citenamefont {Bimberg},\ and\
  \citenamefont {Haisler}}]{Strittmatter2012}%
  \BibitemOpen
  \bibfield  {author} {\bibinfo {author} {\bibfnamefont {A.}~\bibnamefont
  {Strittmatter}}, \bibinfo {author} {\bibfnamefont {A.}~\bibnamefont
  {Schliwa}}, \bibinfo {author} {\bibfnamefont {J.~H.}\ \bibnamefont
  {Schulze}}, \bibinfo {author} {\bibfnamefont {T.~D.}\ \bibnamefont
  {Germann}}, \bibinfo {author} {\bibfnamefont {A.}~\bibnamefont {Dreismann}},
  \bibinfo {author} {\bibfnamefont {O.}~\bibnamefont {Hitzemann}}, \bibinfo
  {author} {\bibfnamefont {E.}~\bibnamefont {Stock}}, \bibinfo {author}
  {\bibfnamefont {I.~A.}\ \bibnamefont {Ostapenko}}, \bibinfo {author}
  {\bibfnamefont {S.}~\bibnamefont {Rodt}}, \bibinfo {author} {\bibfnamefont
  {W.}~\bibnamefont {Unrau}}, \bibinfo {author} {\bibfnamefont {U.~W.}\
  \bibnamefont {Pohl}}, \bibinfo {author} {\bibfnamefont {A.}~\bibnamefont
  {Hoffmann}}, \bibinfo {author} {\bibfnamefont {D.}~\bibnamefont {Bimberg}}, \
  and\ \bibinfo {author} {\bibfnamefont {V.}~\bibnamefont {Haisler}},\ }\href
  {\doibase 10.1063/1.3691251} {\bibfield  {journal} {\bibinfo  {journal}
  {Applied Physics Letters}\ }\textbf {\bibinfo {volume} {100}},\ \bibinfo
  {pages} {093111} (\bibinfo {year} {2012}{\natexlab{b}})}\BibitemShut
  {NoStop}%
\bibitem [{\citenamefont {Kuhlmann}\ \emph {et~al.}(2015)\citenamefont
  {Kuhlmann}, \citenamefont {Prechtel}, \citenamefont {Houel}, \citenamefont
  {Ludwig}, \citenamefont {Reuter}, \citenamefont {Wieck},\ and\ \citenamefont
  {Warburton}}]{Kuhlmann2015}%
  \BibitemOpen
  \bibfield  {author} {\bibinfo {author} {\bibfnamefont {A.~V.}\ \bibnamefont
  {Kuhlmann}}, \bibinfo {author} {\bibfnamefont {J.~H.}\ \bibnamefont
  {Prechtel}}, \bibinfo {author} {\bibfnamefont {J.}~\bibnamefont {Houel}},
  \bibinfo {author} {\bibfnamefont {A.}~\bibnamefont {Ludwig}}, \bibinfo
  {author} {\bibfnamefont {D.}~\bibnamefont {Reuter}}, \bibinfo {author}
  {\bibfnamefont {A.~D.}\ \bibnamefont {Wieck}}, \ and\ \bibinfo {author}
  {\bibfnamefont {R.~J.}\ \bibnamefont {Warburton}},\ }\href {\doibase
  10.1038/ncomms9204} {\bibfield  {journal} {\bibinfo  {journal} {Nature
  Communications}\ }\textbf {\bibinfo {volume} {6}},\ \bibinfo {pages} {8204}
  (\bibinfo {year} {2015})}\BibitemShut {NoStop}%
\bibitem [{\citenamefont {Wang}\ \emph {et~al.}(2016)\citenamefont {Wang},
  \citenamefont {Duan}, \citenamefont {Li}, \citenamefont {Chen}, \citenamefont
  {Li}, \citenamefont {He}, \citenamefont {Chen}, \citenamefont {He},
  \citenamefont {Ding}, \citenamefont {Peng}, \citenamefont {Schneider},
  \citenamefont {Kamp}, \citenamefont {H\"{o}fling}, \citenamefont {Lu},\ and\
  \citenamefont {Pan}}]{Wang2016}%
  \BibitemOpen
  \bibfield  {author} {\bibinfo {author} {\bibfnamefont {H.}~\bibnamefont
  {Wang}}, \bibinfo {author} {\bibfnamefont {Z.~C.}\ \bibnamefont {Duan}},
  \bibinfo {author} {\bibfnamefont {Y.~H.}\ \bibnamefont {Li}}, \bibinfo
  {author} {\bibfnamefont {S.}~\bibnamefont {Chen}}, \bibinfo {author}
  {\bibfnamefont {J.~P.}\ \bibnamefont {Li}}, \bibinfo {author} {\bibfnamefont
  {Y.~M.}\ \bibnamefont {He}}, \bibinfo {author} {\bibfnamefont {M.~C.}\
  \bibnamefont {Chen}}, \bibinfo {author} {\bibfnamefont {Y.}~\bibnamefont
  {He}}, \bibinfo {author} {\bibfnamefont {X.}~\bibnamefont {Ding}}, \bibinfo
  {author} {\bibfnamefont {C.~Z.}\ \bibnamefont {Peng}}, \bibinfo {author}
  {\bibfnamefont {C.}~\bibnamefont {Schneider}}, \bibinfo {author}
  {\bibfnamefont {M.}~\bibnamefont {Kamp}}, \bibinfo {author} {\bibfnamefont
  {S.}~\bibnamefont {H\"{o}fling}}, \bibinfo {author} {\bibfnamefont {C.~Y.}\
  \bibnamefont {Lu}}, \ and\ \bibinfo {author} {\bibfnamefont {J.~W.}\
  \bibnamefont {Pan}},\ }\href@noop {} {\bibfield  {journal} {\bibinfo
  {journal} {Physical Review Letters}\ }\textbf {\bibinfo {volume} {116}},\
  \bibinfo {pages} {213601} (\bibinfo {year} {2016})}\BibitemShut {NoStop}%
\bibitem [{\citenamefont {Trotta}\ \emph {et~al.}(2014)\citenamefont {Trotta},
  \citenamefont {Wildmann}, \citenamefont {Zallo}, \citenamefont {Schmidt},\
  and\ \citenamefont {Rastelli}}]{Trotta2014a}%
  \BibitemOpen
  \bibfield  {author} {\bibinfo {author} {\bibfnamefont {R.}~\bibnamefont
  {Trotta}}, \bibinfo {author} {\bibfnamefont {J.~S.}\ \bibnamefont
  {Wildmann}}, \bibinfo {author} {\bibfnamefont {E.}~\bibnamefont {Zallo}},
  \bibinfo {author} {\bibfnamefont {O.~G.}\ \bibnamefont {Schmidt}}, \ and\
  \bibinfo {author} {\bibfnamefont {A.}~\bibnamefont {Rastelli}},\ }\href
  {\doibase 10.1021/nl500968k} {\bibfield  {journal} {\bibinfo  {journal} {Nano
  Letters}\ }\textbf {\bibinfo {volume} {14}},\ \bibinfo {pages} {3439}
  (\bibinfo {year} {2014})}\BibitemShut {NoStop}%
\bibitem [{\citenamefont {Wei}\ \emph {et~al.}(2014)\citenamefont {Wei},
  \citenamefont {He}, \citenamefont {He}, \citenamefont {Lu}, \citenamefont
  {Pan}, \citenamefont {Schneider}, \citenamefont {Kamp}, \citenamefont
  {H{\"{o}}fling}, \citenamefont {McCutcheon},\ and\ \citenamefont
  {Nazir}}]{Wei2014}%
  \BibitemOpen
  \bibfield  {author} {\bibinfo {author} {\bibfnamefont {Y.-J.}\ \bibnamefont
  {Wei}}, \bibinfo {author} {\bibfnamefont {Y.}~\bibnamefont {He}}, \bibinfo
  {author} {\bibfnamefont {Y.-M.}\ \bibnamefont {He}}, \bibinfo {author}
  {\bibfnamefont {C.-Y.}\ \bibnamefont {Lu}}, \bibinfo {author} {\bibfnamefont
  {J.-W.}\ \bibnamefont {Pan}}, \bibinfo {author} {\bibfnamefont
  {C.}~\bibnamefont {Schneider}}, \bibinfo {author} {\bibfnamefont
  {M.}~\bibnamefont {Kamp}}, \bibinfo {author} {\bibfnamefont {S.}~\bibnamefont
  {H{\"{o}}fling}}, \bibinfo {author} {\bibfnamefont {D.~P.}\ \bibnamefont
  {McCutcheon}}, \ and\ \bibinfo {author} {\bibfnamefont {A.}~\bibnamefont
  {Nazir}},\ }\href {\doibase 10.1103/PhysRevLett.113.097401} {\bibfield
  {journal} {\bibinfo  {journal} {Physical Review Letters}\ }\textbf {\bibinfo
  {volume} {113}},\ \bibinfo {pages} {097401} (\bibinfo {year}
  {2014})}\BibitemShut {NoStop}%
\bibitem [{\citenamefont {Ulrich}\ \emph {et~al.}(2011)\citenamefont {Ulrich},
  \citenamefont {Ates}, \citenamefont {Reitzenstein}, \citenamefont
  {L{\"{o}}ffler}, \citenamefont {Forchel},\ and\ \citenamefont
  {Michler}}]{Ulrich2011}%
  \BibitemOpen
  \bibfield  {author} {\bibinfo {author} {\bibfnamefont {S.~M.}\ \bibnamefont
  {Ulrich}}, \bibinfo {author} {\bibfnamefont {S.}~\bibnamefont {Ates}},
  \bibinfo {author} {\bibfnamefont {S.}~\bibnamefont {Reitzenstein}}, \bibinfo
  {author} {\bibfnamefont {A.}~\bibnamefont {L{\"{o}}ffler}}, \bibinfo {author}
  {\bibfnamefont {A.}~\bibnamefont {Forchel}}, \ and\ \bibinfo {author}
  {\bibfnamefont {P.}~\bibnamefont {Michler}},\ }\href {\doibase
  10.1103/PhysRevLett.106.247402} {\bibfield  {journal} {\bibinfo  {journal}
  {Physical Review Letters}\ }\textbf {\bibinfo {volume} {106}},\ \bibinfo
  {pages} {247402} (\bibinfo {year} {2011})}\BibitemShut {NoStop}%
\bibitem [{\citenamefont {F{\"{o}}rstner}\ \emph {et~al.}(2003)\citenamefont
  {F{\"{o}}rstner}, \citenamefont {Weber}, \citenamefont {Danckwerts},\ and\
  \citenamefont {Knorr}}]{Forstner2003}%
  \BibitemOpen
  \bibfield  {author} {\bibinfo {author} {\bibfnamefont {J.}~\bibnamefont
  {F{\"{o}}rstner}}, \bibinfo {author} {\bibfnamefont {C.}~\bibnamefont
  {Weber}}, \bibinfo {author} {\bibfnamefont {J.}~\bibnamefont {Danckwerts}}, \
  and\ \bibinfo {author} {\bibfnamefont {a.}~\bibnamefont {Knorr}},\ }\href
  {\doibase 10.1103/PhysRevLett.91.127401} {\bibfield  {journal} {\bibinfo
  {journal} {Physical Review Letters}\ }\textbf {\bibinfo {volume} {91}},\
  \bibinfo {pages} {127401} (\bibinfo {year} {2003})}\BibitemShut {NoStop}%
\bibitem [{\citenamefont {Ramsay}\ \emph {et~al.}(2010)\citenamefont {Ramsay},
  \citenamefont {Gopal}, \citenamefont {Gauger}, \citenamefont {Nazir},
  \citenamefont {Lovett}, \citenamefont {Fox},\ and\ \citenamefont
  {Skolnick}}]{Ramsay2010}%
  \BibitemOpen
  \bibfield  {author} {\bibinfo {author} {\bibfnamefont {A.~J.}\ \bibnamefont
  {Ramsay}}, \bibinfo {author} {\bibfnamefont {A.~V.}\ \bibnamefont {Gopal}},
  \bibinfo {author} {\bibfnamefont {E.~M.}\ \bibnamefont {Gauger}}, \bibinfo
  {author} {\bibfnamefont {A.}~\bibnamefont {Nazir}}, \bibinfo {author}
  {\bibfnamefont {B.~W.}\ \bibnamefont {Lovett}}, \bibinfo {author}
  {\bibfnamefont {A.~M.}\ \bibnamefont {Fox}}, \ and\ \bibinfo {author}
  {\bibfnamefont {M.~S.}\ \bibnamefont {Skolnick}},\ }\href {\doibase
  10.1103/PhysRevLett.104.017402} {\bibfield  {journal} {\bibinfo  {journal}
  {Physical Review Letters}\ }\textbf {\bibinfo {volume} {104}},\ \bibinfo
  {pages} {017402} (\bibinfo {year} {2010})}\BibitemShut {NoStop}%
\end{thebibliography}

%

\end{document}